# Timing Channel in IaaS: How to Identify and Investigate

Xiao Fu, Rui Yang, Xiaojiang Du, Bin Luo

**Abstract**—Recently, the IaaS (Infrastructure as a Service) Cloud (e.g., Amazon EC2) has been widely used by many organizations. However, some IaaS security issues create serious threats to its users. A typical issue is the timing channel. This kind of channel can be a cross-VM information channel, as proven by many researchers. Because it is covert and traceless, the traditional identification methods cannot build an accurate analysis model and obtain a compromised result. We investigated the underlying behavior of the timing channel from the perspective of the memory activity records and summarized the signature of the timing channel in the underlying memory activities. An identification method based on long-term behavior signatures was proposed. We proposed a complete set of forensics steps including evidence extraction, identification, record reserve, and evidence reports. We studied four typical timing channels, and the experiments showed that these channels can be detected and investigated, even with disturbances from normal processes.

## 1. INTRODUCTION

Recently, there has been significant attention drawn to the security of the Infrastructure as a Service (IaaS) cloud [1], since a compromised cloud may pose a threat to multiple cloud-backend application domains [2, 3]. A particular focus is on the timing channel [4, 5]. Ristenpart et al. [6] first proposed cross-VM timing channels in the cloud environment. They stated that any physical machine resources multiplexed between the attacker and the target might form a potential leakage channel between the virtual machines. Recent studies show that this kind of attack can successfully steal from co-resident VM instances some private information, including private keys, which are set up by key management schemes [7, 8, 9].

To prevent the timing channel, some protection mechanisms have been proposed. sHype [10] is a mandatory access control (MAC)-based security extension to the Xen hypervisor that allows the application of various security policies on VMs. Lares [11] presented a hybrid approach, giving security tools the ability to actively monitor while benefitting from the increased security of an isolated virtual machine. They all focus on providing an integrity measurement to the hypervisor by introducing a software component.

However, these protection mechanisms may fail because the timing channel is created by sharing resources. These mechanisms observe behaviors of a VM by monitoring hypercalls in a hypervisor and look for the regularity in the system operation records. The system operations records obtained are an abstracted behavior of the system layer. The information they contain is not complete. Timing channels use time information from shared resources, which are not reflected by the general system operation records. In this paper, we propose a behavior observation mode on memory access activity. The lower memory activity records, which provide crucial evidence for forensics and a good basis for further analysis, can also provide complete information to understand the behavior of the timing channel process.

The presence of a timing channel is difficult to identify. It can run continuously without leaving a trace or an alarm. All behaviors on the timing channel are legal and normal. This creates challenges to their identification and forensics. After careful investigation, we found that, although single behavior is normal, the timing channel shows regular patterns in long-term behaviors. These signatures provide a good basis for our automatic detection and forensic feature.

In this paper, only cross-VM timing channels new to IaaS are considered. Four typical scenarios (e.g. CPU load based, Cache based, shared memory based channels, IP timing channel) are demonstrated, and their features are studied in-depth. We present an identification method in IaaS and introduce an investigation on the timing channel. Our method includes three steps. First, to record the behavior of the timing channel, a page-level memory record and packet record are designed. Second, an automatic identification algorithm is proposed based on the timing channel's long-term behavior. Finally, a memory dump will be obtained, and the binary code of suspicious processes will be analyzed for the forensics scheme.

A prototype system was implemented on Xen. The effectiveness of identifying the four typical timing channels was investigated. Meanwhile, we attempted to identify a cache-based side-channel attack and simulate a flush-reload [12] attack to extract the private encryption keys from GnuPG. The experiment results show that our system can identify these channels, even with the disturbances from normal processes, and the introduced overhead is up to 42% in the worst case, which is lower compared to current methods.

This paper makes the following contributions:

- We introduce memory activity records, which provide a good basis for the timing channel and can extract meaningful information. The time information of the memory record reflects the behavior of the timing channel.
- We summarize the behavior signature of the timing channel based on the investigation of long-term behavior activities on the memory activity. We build a precise analysis model and design an automatic identification algorithm.
- We implemented a complete set of forensic tools. Based on the recognition function, we add the evidence acquisition reservation and analysis, which achieves the forensic schema for the timing channel in the cloud.

- Xiao Fu is with the State Key Laboratory for Novel Software Technology, Nanjing University, China. E-mail: fuxiao@nju.edu.cn.
- Rui Yang is with the State Key Laboratory for Novel Software Technology, Nanjing University, China. E-mail: mg1432016@smail.nju.edu.cn.
- Xiaojiang Du is with the Dept. of Computer and Information Sciences, Temple University, USA. Email: dxj@ieee.org
- Bin Luo is with the State Key Laboratory for Novel Software Technology, Nanjing University, China. E-mail: luobin@nju.edu.cn.

- We built our prototype system on the Xen platform and tested it with four typical time channels. Additionally, we compared identification algorithms from similar works in our experiments and introduced an analysis method for memory activity records from packet records on the network timing channel.

The rest of this paper is organized as follows: Section 2 discusses the related work. Section 3 describes the timing channel threat in the cloud. Section 4 describes the long-term identification. In Section 5, an overview of the prototype system is given. Section 6 describes the experiment and evaluation. Our paper concludes in Section 7.

## 2. RELATED WORK

Recently, several efforts have been made. A typical proposal is a hardware-based solution [13]. However, these solutions have high costs and latency. Considering this, [5] HomeAlone was designed to proactively detect the co-residence of unfriendly VMs to offer immediately deployable protections to the IaaS. It detected the presence of a malicious VM by acting as a timing channel receiver and observing the cache timing anomalies caused by another receiver's activities. Different from HomeAlone, XenPump [14] placed hooks into the Xen hypervisor monitored hypercalls and added latencies to mitigate the threat from the timing channels.

Currently, the main detection method focuses on the timing channel behavior. For instance, C2Hunter [15] presented a two-phase synthesis detection algorithm using Markov and Bayesian models based on error-corrected four-state automation, which modeled the timing channel scenario in cloud computing. [16] used a directed information flow graph, taking advantage of the source code analysis to detect the timing channel.

In [17], the authors illustrated the threat from IP covert timing channels and developed two methods for detecting IP covert timing channels by identifying the regularity of the inter-transmission times. These methods worked well for noiseless channels but failed with a high level of noise.

Compared to HomeAlone, our method not only detects the existence of malicious behavior but also collects related evidence of these timing channels. Several papers [18,19,20] have studied related security issues. In addition, different from XenPump, our method does not insert additional modules into the hypervisor. We use the EPT (Extended Page Table) modification interface [21] supplied by Xen in dom0 to monitor the memory activity and place a hook on the NIC to retrieve packet information. Compared with the C2Hunter [15] and [16], our method focuses on the underlying view of single process memory activities, which have more specific and useful information. This not only allows our method to have a higher detection rate but also significantly reduces the false positives rate.

Our work concentrates on investigating (including evidence collection and analysis) timing channels in IaaS. Research on how to investigate them after identification is necessary. We have not found any other work on this issue.

## 3. BACKGROUND AND MOTIVATION

### 3.1 Timing channel in IaaS

IaaS refers to online services that abstract the user from the details of the infrastructure such as physical computing resources, location, data partitioning, scaling, security, backup, etc. Although virtualization technology provides strong isolation for the cloud, the service provider is a shared infrastructure. One of the problems introduced by a shared environment to IaaS is the timing channel. The main idea of the timing channel is that every logical operation in a computer takes time to execute, and the time can differ based on the shared environment in the IaaS. With precise measurements of the time for each operation, an attacker can work backward to the shared environment [5]. The most convenient resources are a CPU, cache, memory, and a network, which makes the timing channel more effective.

In this paper, only the cache-based, load-based, and shared memory belonging to cross-VM channels are considered. We classified them into local channel categories.

***Local channel***: Malicious processes $P_i$ and $P_j$ are located on the same hardware platform. $P_i$ leaks confidential information to the $P_j$ using the timing channel through shared local resources (e.g., cache, shared memory, CPU load.). They encode messages into the time information of the shared source such as cache access latency, memory access latency, etc. [4, 22, 23].

***Network channel***: Malicious processes *Pk* and *Px* communicate with each other through the network resource. The confidential information is encoded into the different packet time intervals. They are located on different domains or in different hardware platforms, which makes it easier to implement and create more significant threats (e.g., IP timing channel) [17].

### 3.2 Threat scenarios of timing channel

**CPU load-based channels.** [23] These can be approximated using the amount of time taken for certain computations. The confidential information is pre-encoded into a binary sequence. The sender and receiver transfer information by changing and observing the CPU load according to a certain communication protocol. For example, a long waiting time to complete a task transmits bit 1, otherwise, bit 0 is transmitted.

**Cache-based channels.** [4] The cache-based channel considers the different cache access latencies as different bits. The sender uses the idle cache access as transmitting bit 0 and the frequent accesses to a memory block as transmitting bit 1. The receiver accesses a memory block and observes the access latencies. High latency indicates the sender is removing the receiver's data from the cache, and bit 1 is transmitted; otherwise, bit 0 is transmitted.

**Shared memory-based channels.** [22] The shared memory-based channel considers different memory access intervals as different bits. The sender sends covert messages by controlling the data sending time, and the receiver receives the message by observing the data arrival time. The confidential information is encoded into the different intervals. For example, bit 1 and bit 0 denote longer and shorter intervals.

**IP timing channel.** [17] The receiver and sender agree on a time interval and the starting protocol. During each time interval, the sender either transmits a single packet or maintains silence. The receiver and sender also agree on a different time interval. The sender transmits a single packet with a different time interval. The receiver monitors the interval between adjacent packets to decode the message.

### 3.3 Challenges

In recent studies, the timing channel has been modeled using transitional behavior analysis methods, such as system operation analysis and tainted analysis. However, it is difficult to construct an accurate analysis model because of its misinterpretation of the behavior semantics, which reduces the recognition rate and increases the false alarm rate. Meanwhile, the isolation mechanism of the cloud makes forensics more difficult and exposes the existence of monitoring procedures, leading to an anti-forensic response. To identify and investigate the timing channel in the cloud, we need solve the following challenges:

- **How to extract meaningful information.** The timing channel is concerned with the timing information. All of its behaviors are ordinary user behavior, such as reading files, loading programs, etc., which makes it difficult for the traditional behavior log records to correctly express their real purpose. To obtain valuable information, we must find another way to express the timing channel behavior information. This paper presents a method based on memory activity records analysis to express the processes of the timing channel.

- **There is still no accurate analysis model for the local channel.** In the existing research, the local channel is analyzed using the system behavior or the information flow. Due to the concealment of the time information, the analysis model constructed from the traditional behavior analysis makes accurate identification difficult and leads to a high false alarm rate. To solve this, we summarize the long-term behavior signature of the timing channel based on the memory activity records and build an accurate model with a high recognition rate.

- **How to investigate in the cloud environment under isolated conditions without being discovered by the customer.** The isolation of the cloud environment makes it necessary to modify the client or modify the kernel when monitoring the client. This leaves traces, making it easy for malicious programs to discover and create an anti-forensic response. Therefore, the main body of the monitor is placed in a virtual machine manager, and only the existing system interface is called to implement the monitoring of the memory record.

# 4. LONG-TERM BEHAVIOR SIGNATURE BASED IDENTIFICATION METHOD

## 4.1 Timing channel characters

As shown in figure 1, during a typical transmission cycle of the timing channel, the confidential information is transmitted from the sender to the receiver located on different virtual machines. The sender of the timing channel encodes the information into binary bits and changes the properties of the shared resources according to the bits. The receiver observes the changes and decodes the confidential information from these changes. The sender and receiver predetermine the parameters and repeat the cycle until all the confidential information has been transmitted.

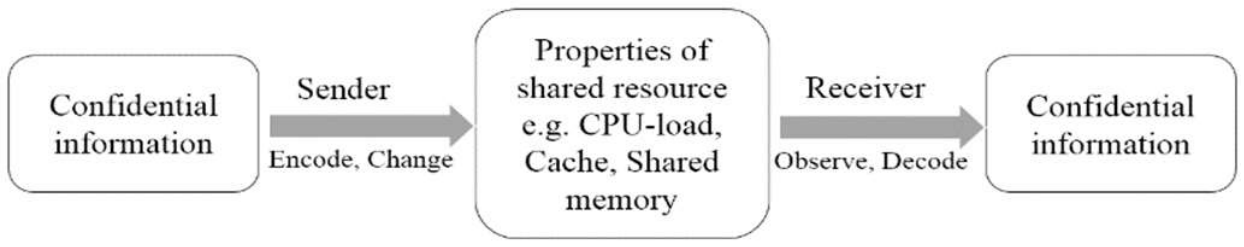

**Figure1. Transmission procedure of Timing channel**

Summarizing the above describes the long-term behavior signature of the timing channel: **Repetitive access to a single shared resource over a period of time.**

There are two ways for the timing channel to encode information: **Storage timing channel** (STC): in the same time interval, the sender's selection is active or silent. Activity represents one and silence is zero. **Distinct timing channel** (DTC): different time intervals for encoding information. For example, a long interval is one, and a short interval is zero, which can be seen on the network channel.

During behavioral analysis, we use the network packet communication situation on the network channel to reflect the timing channel behavior. For the local channel, we acquire system operations from domU by adding a module into the hypervisor to intercept hypercalls, but additional changes to the hypervisor are required. Additionally, the system operation information, such as the time information, is not complete and cannot be accessed directly. To get complete information about the local channels, we investigate the memory access activity.

The CPU accesses the memory to obtain data when it executes instructions or extracts data. There are three permissions for accessing memory. The execution permission is used for the instruction area of the processed memory area. For the data area, read and write permission are granted. After careful investigation, the transmission cycle of the timing channel is repeatedly executed to complete the data transmission. Every transmission cycle accesses the same area of instruction and data. Therefore, we observed the lower behavior to determine the regularity of the timing channel. Additionally, the probability of repeatedly executing the same instruction area is higher than the data area, so we focused on the memory data read operation to investigate timing channel characteristics.

In conclusion, we identified the following signatures of the local channel in IaaS:

Ⅰ、 During the running of the timing channel which is always fixed to the same process.

Ⅱ、 During the running of the local channel, the process will repeatedly access the fixed memory page at intervals. For the network channel, the packet will repeatedly be sent at intervals or received from a fixed IP and port to a fixed IP and port.

Ⅲ、 The **DTC** behavior logs show the level of two different fixed intervals.

Ⅳ、 The **STC** behavior logs show a variety of time intervals, and the maximum interval time should be an integer multiple of the minimum interval time.

## 4.2 Detection method

We designed an identification method based on the long-term behavior signature, despite the different records of three types of local channels. The input for our method is the interval time of the adjacent activity. Therefore, we pre-computed the interval time of the memory activity records. As shown in Algorithm 1, our method has the following phases:

**Algorithm 1** identify timing channel

```
1:  Input: the memory activity records and network packet records
2:  Output: the suspicious process
3:  if accessd(x) >= RepeatThreshold
4:     for I = 1 to accessd(x)
5:        if pid[0](page(x)) != pid[i](x)
6:           break
7:     if i == accessd(x)
8:        suspicious.add(x)
9:  for i = 1 to accessd(x)
10:    interval.add(time[i](x) – time[i-1](x))
11: SORT(interval)
12: for i = 1 to accessd(x)
13:    if (|interval[i] − interval[i-1]|/ interval[i-1]) <= K1
14:       group.add(interval[i])
15:    else
16:       group.addGroup()
17:       group.add(interval[i])
18: for i = 1 to length(group)
19:    b = line_program(group[i])
20:    if b >= K2
21:       suspicious.remove(x)
22:       break
23: if length(group) == 2
24:       type = DTC
25: else if length(group) > 2 and
          average(group[n]) % average(group[1]) == 0
26:       type = STC
```

**Found repeat operation**: for the memory activity records, we first locate the memory page that was repeatedly accessed. In addition to the process of accessing, the repeated page should also be fixed. For packet records, the same IP and the same port that repeatedly sends out or receives incoming packets is targeted, and the destination IP and packets port should be fixed. As shown in lines 3 to 8, based on signatures I and II, we perform initial filtering to remove the meaningless records. Accessd() represents the access records of the memory page x. pid() represents the process of accessing the memory page x. RepeatThreshold represents the minimum number of accesses to the memory to determine the suspicious process. The default is set to 100. After the initial filter, the list of suspicious processes is added for the next analysis.

**Compute interval**: We extracted and sorted the records of the previous suspicious process, then computed the interval time of the adjacent operations, as shown in lines 9 to 10.

**Sorted IA(inter-arrival) times list:** we sort the interval time of the adjacent operations, as shown in line 11.

**Smoothness calculation:** from the sorted list, we set the time thresholds, and we separate the similar IA times into each threshold. Then, we compute the relative difference in each threshold. We use line programming to compute the smoothness. For the timing channel, the regression coefficient should be close to zero in each threshold. For the timing channels, the majority of the pairwise differences in the sorted list of IA times will be very small. It is only large for jumps in the step function, as shown in lines 12 to 22. We calculate the difference between adjacent thresholds. When the adjacent value difference is greater than K1, a new threshold is created. If the adjacent value difference is less than K1, the value is added. Finally, to calculate the regression coefficient b for each group, when b is less than K2, the data group is smooth and effective. K1 and K2 are set to 0.1 and 0.01 by default.

**Pattern match:** as shown in lines 23 to 26, we identify several thresholds, based on signatures III and IV. The DTC will show two smoothness thresholds. The STC will show more than two smoothness thresholds, and the maximum threshold should be an integer multiple of the minimum threshold.

## 5. IMPLEMENTATION

The introduction of the memory activity record solves the challenge of finding meaningful information that can record the timing channel. Adding the recognition algorithm proposed above, we obtain the evidence acquisition and recognition function in the forensic process. Therefore, we add the evidence reservation and analysis module, which forms a complete set of forensics tools to achieve the forensics for the timing channel in the cloud environment.

We implemented our prototype system on a desktop computer with an Intel® Core™ i5-3330 3.00 GHz CPU, 8 GB RAM, and 256 KB L2 cache. The version of Xen hypervisor was 4.4.3.

Our system included a monitor, detector, evidence collector, and verifier module. The monitor collected the system operation records. The detector identified the suspicious processes on the timing channel from the records. The evidence controller collected the relevant evidence and stored it in the database. The verifier obtained the binary code of the suspicious process and analyzed the communication protocol to reconstruct the crime scene.

### 5.1 Monitor and Detector

We placed the monitor modules into dom0, which was transparent to domU. The monitor module was responsible for recording the memory activity.

For the memory activity recording, we used Intel VT EPT [21] techniques to monitor the guest VM's memory region. The monitoring information was delivered to dom0 using event channels provided by Xen. No modifications of the guest VMs were needed in the procedure. The monitor was first initialized using the *vmi_init* function. Then, the *vmi_register_event* function registered the memory events on the guest VM's memory region to monitor the VM's memory in real-time. To trigger the event, the access control mechanism in EPT was adopted to set the read permission on the target memory region. When a guest process wanted to access the memory page, a violation event was triggered and trapped in the hypervisor. Meanwhile, the access control was canceled to let the guest process continue executing. The violation event was passed through the event channel in the Xen hypervisor, then the *mem_monitor_cb* function was called to handle it. Function *mem_monitor_cb* recorded crucial details such as PID, cr3, process name, access time, page, etc. After handling the violation event, the *vmi_register_event* function was invoked again to register the memory event to reset the access permission on that page and to wait for the next memory access request.

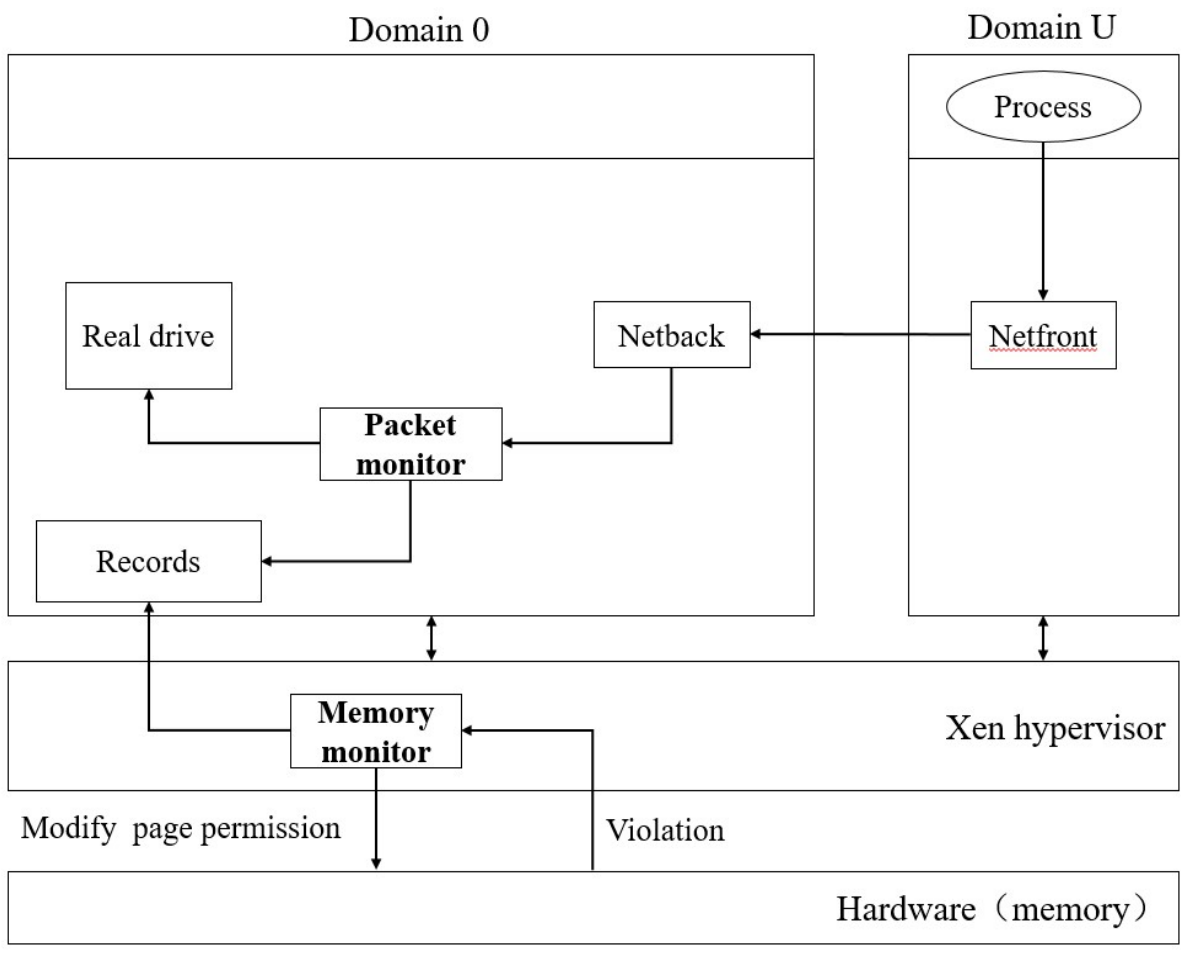

**Figure 2. The architecture of the monitor**

Packet recordings were performed as follows. When the process sent a packet, the guest operating system wrote it to a virtual network interface card. We obtained the packet information using the *libpcap* tool. The structure *pcap_pkthdr* provided by the *libpcap* tool obtained accurate time information. When the *libpacap* received a packet and passed the packet to the monitor, the monitor recorded the packet information, including the source IP, source port, time, destination IP, destination port, etc.

We placed the detector module in the dom0 which obtained records from the database to analyze. The outputs of the detector were the suspicious processes. When detecting the presence of the timing channel, the detector notified the evidence collector, while the next detection cycle continued probing the timing channel data and combining the data belonging to the same timing channel.

## 5.2 Evidence Collector and Verifier

When the evidence collector received the timing channel signal from the detector, it used the VMM introspection mechanism to obtain the guest memory dump through the command dump-core. The memory dump contained the evidence of the timing channel's suspicious processes.

The verifier was implemented as a volatility [24] plugin, which is a memory extraction utility framework. If the system detected the network channel, we extracted the network connection through the linux_netstat plugin, then identified the suspicious process using the corresponding suspicious port. After that, we began our memory analysis. We identified the suspicious processes' memory address through the linux_proc_maps plugin. The linux_dump_map copied the data of the suspicious process.

After obtaining the binary code from the suspicious process, the code was disassembled using the objdump tool. Then, we performed the static analysis based on the timing channel code signature. This step further confirmed the presence of the timing channel to reduce the false positives and obtain the communication protocol from the analysis of the communication code.

In our prototype system, we stored the record in each step and traced the source of the evidence during the subsequent investigation. The entire system contained three databases to store the evidence: the monitor database, the detector database and the memory evidence database. The monitor database kept all of the records during monitoring. The detector database stored the data filtered by the identification and joined the suspicious process info in the original records. The memory evidence database held each suspicious process memory dump and the results of the binary code analysis. The records from the three databases were connected to form a complete chain of evidence to ensure the validity and integrity.

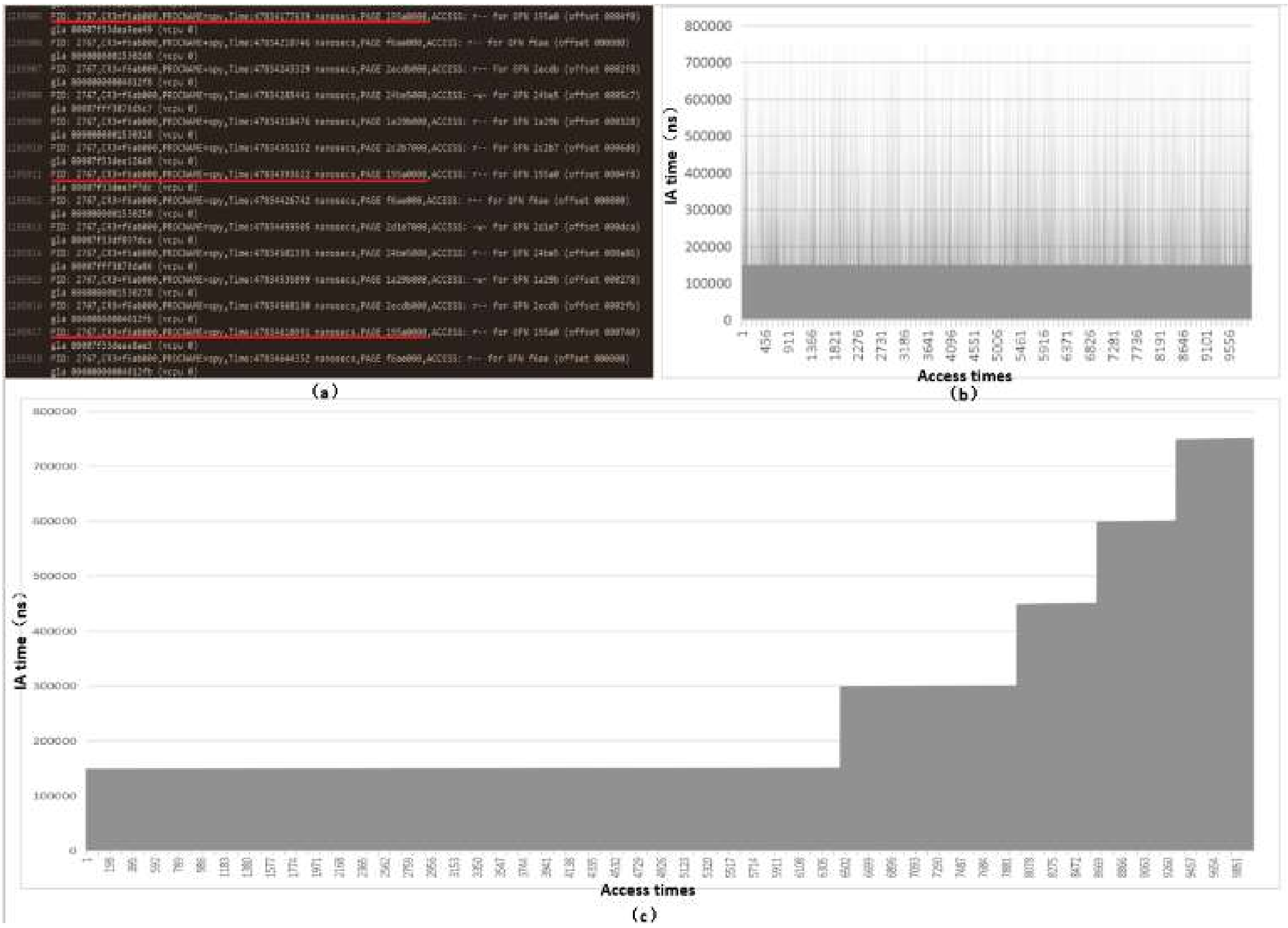

**Figure 3. Experiment on the local channel. (a) Memory activity records. (b) The IA records of memory page 0x195a0000. (c) The sorted IA.**

## 6. EVALUATION

We tested the four typical local timing channels in our system. We ran Ubuntu14.04 in dom0 and two guest VMs, each of which was allocated 1024 MB of virtual memory.

### 6.1 Case Study for the Local channel

Considering the cache-based timing channel, we used the flush-reload code to simulate an attack, which used the cache-based channels to extract the private encryption keys from a victim program running GnuPG. We ran the flush-reload spy program on domA and ran the GnuPG encryption program on domB. domA had the GnuPG encryption table address in domB and mapped to their memory address space.

**Step 1**. As shown in Figure 3(a), we checked the monitoring records in domA. Then, we found the page 0x195a0000, which had repeatedly been accessed over a period of time. We marked it and checked the access. After comparison, we found that the process of accessing the page was always the PID for 2767 belonging to spy.exe.

**Step 2**. As shown in Figure 3(b), we filtered out the accessing memory pages records 0x195a0000 and calculated the time interval between the two adjacent access records.

**Step 3**. As shown in Figure 3(c), with the sorted IA list of the memory page 0x195a0000 access time, we filtered the meaningless thresholds, which had a count less than 5% of a total. In this case, we identified five thresholds.

**Step 4**. Smoothness calculation. A meaningful threshold for the timing channel is smoothness. With the linear programming technique, the regression coefficient should be close to zero. We calculated the five thresholds. All the results were close to zero.

**Step 5.** Pattern match. By comparing the regularity, we found its behavior pattern matched the STC pattern. That was the multiple integer relationship between the maximum threshold and the minimum threshold. In our records, the minimum interval was 200000 ns.

**Step 6**. After the timing channel was found, the evidence collector stored the memory dump of domA and domB.

The additional information was used to determine the type of timing channel. We identified the suspicious process spy, and we also found the process of the GnuPG encryption program. The page activities were cross-performed, and the domA and domB shared a physical cache. Thus, it was determined that the currently running timing channel was a cache-based channel attack and its target was the GnuPG encryption process in domA.

In conclusion, there was a timing channel attack in the domB, and we conducted a binary code analysis to confirm its existence with more evidence.

When the suspicious process was detected, the system automatically extracted the current memory dump after the suspicious process was confirmed. We located the memory address of the process based on the PID record and then extracted data from the target memory for further evidence analysis.

### 6.2 Case Study for the Network channel

In this experiment, we simulated a simple IP timing channel. We set the different interval times (0.3 sec and 0.5 sec) for the entire communication. We ran the sender process on domA and the receiver on domB. An empty content UDP packet was transported between them.

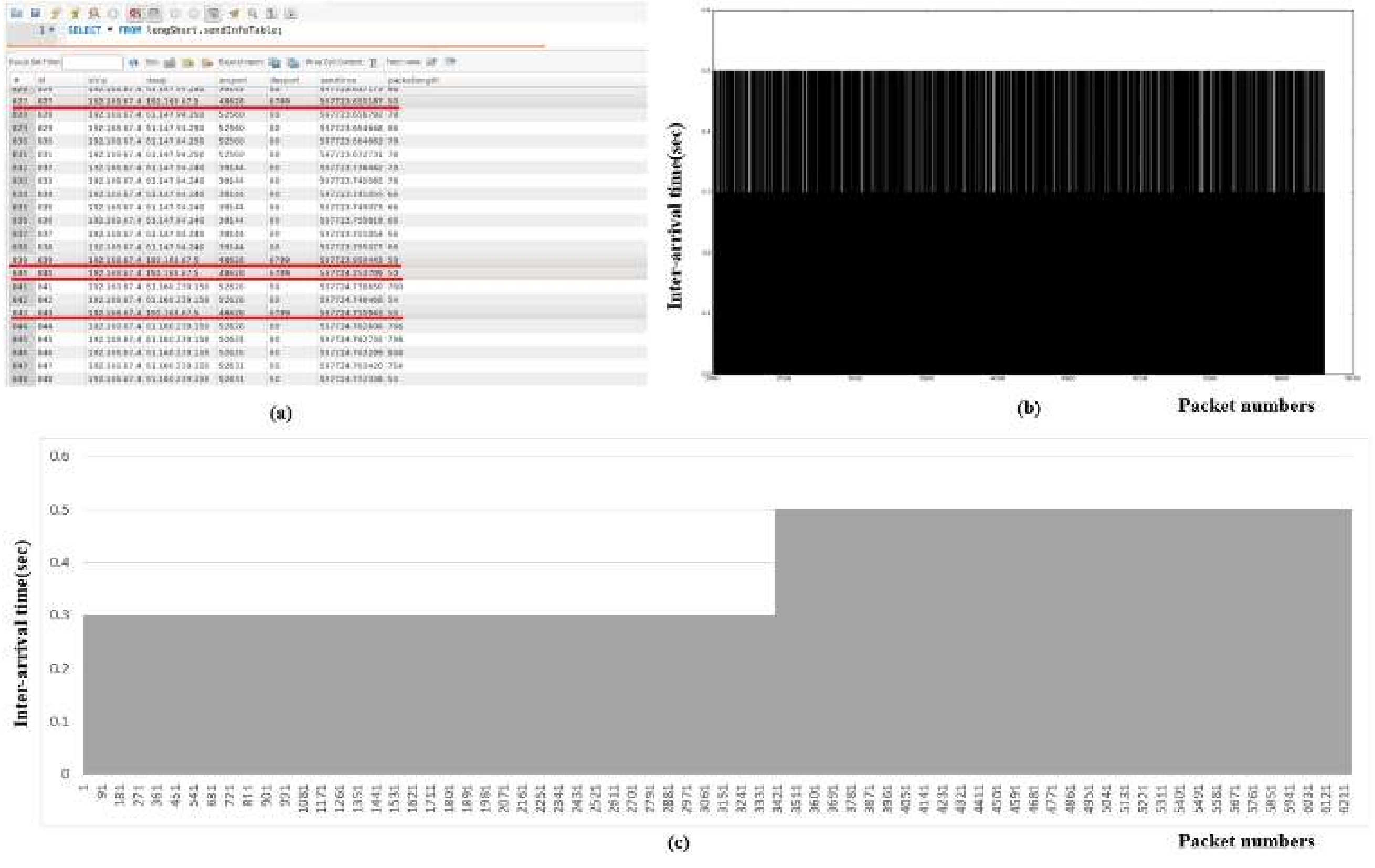

**Figure4. experiment of network channel. (a) Sending packets records. (b) The packet interval pattern on port 48628. (c) The IA records of memory page 0x195a0000.**

**Step 1**. We reviewed the packet records and found packets repeatedly sending on port 48628 in domA. Then we determined whether its destination IP and port were fixed. We found all the packets sent to the IP belonged to domB on port 6789.

**Step 2**. We filtered out the sending packets records by port 48628 and calculated the time interval between two adjacent packet records.

**Step 3**. By comparing the regular packet sending interval, we sorted the packet records IA and identified two meaningful thresholds.

**Step 4**. Smoothness calculation. We calculated the two thresholds using the line program. All the regression coefficients were close to zero.

**Step 5**. Pattern match. We identified two different interval thresholds; the high was 0.5 sec, and the low was 0.3 sec. This is a typical DTC pattern.

**Step 6**. We located the suspicious process iptming.exe on port 48628, then marked it and copied its process data. We disassembled the iptiming.exe binary code using the objdump tool and located the communication code. We found that its communication protocol was the DTC protocol. The time intervals were 0.3 seconds and 0.5 seconds. We defined the maximum interval to send 1 and the log interval to send 0. The IP timing channel communication protocol was restored, and the monitored records were consistent with the case.

### 6.3 Effectiveness

We introduced a similar approach to the study to evaluate the efficiency and performance of the proposed method. The similar methods in the current field use the two-phase synthesis detection algorithm presented by C2Hunter [15]. We also introduced the packet records detection method from network timing channel research, such as examining patterns in the variance and ϵ-similarity [17], which also uses the memory activity records for the input.

**The two-phase synthesis detection algorithm** is synthesized using the Markov detection algorithm and the Bayesian detection algorithm. If the change pattern of the shared resource properties is closer to the Markov model, the sequence of the operations is transferring confidential information through a timing channel. The normal operation sequences are modeled into the Bayesian model. If an operation sequence deviates from the Bayesian model, a timing channel occurs. Markov and Bayesian models in the two-phase synthesis algorithm are complementary. The Markov detector detects the timing channels, and the Bayesian detector distinguishes the timing channels from the normal operation sequences.

**Examining patterns in the variance:** This method examines whether the variance in the IA remains constant. The traffic is separated into non-overlapping windows. For each window, the standard deviation of the IA times is computed. To compute the heuristic measure of regularity, the pairwise differences between the window for each pair is calculated. Finally, to obtain a summary statistic, the standard deviation of the pairwise differences is computed.

**ϵ-Similarity:** The measure is derived from the sorted IA times. From this sorted list, the relative difference between each pair of consecutive points is computed. Then, a measure of similarity is computed, ϵ-Similarity, by computing the percentage of relative differences that are less than ϵ.

This process resulted in a good detection rate but had more false positives. To test the effectiveness of our method, we set up datasets with different conditions, which were classified into two groups, normal condition, and noise condition. The noise condition included many extra normal processes running beside the timing channel. Each dataset was tested at least 1000 times in our

**Table1. Detection results of the local channel**

| Dataset | Two-phase | | Variance | | ϵ-similarity | | Long-term signature | |
|---|---|---|---|---|---|---|---|---|
| | Success rate (%) | False positive (%) | Success rate (%) | False positive (%) | Success rate (%) | False positive (%) | Success rate (%) | False positive (%) |
| **Normal** | | 1.13 | | 12.31 | | 10.0 | | 1.33 |
| **cache channel** | 93.33 | 1.89 | 76.33 | 13.26 | 91.85 | 12.1 | 99.97 | 1.06 |
| **load based channel** | 95.24 | 4.46 | 72.32 | 15.31 | 87.46 | 13.21 | 99.6 | 1.11 |
| **share memory channel** | 97.57 | 2.53 | 74.27 | 14.89 | 84.94 | 13.89 | 98.59 | 1.21 |
| **IP timinmg channel** | | | 60.52 | 16.55 | 85.10 | 10.50 | 99.50 | 1.66 |
| **Flush-reload** | 93.29 | 2.12 | 75.21 | 13.51 | 90.54 | 12.25 | 99.67 | 2.09 |
| **Cache channel with noise** | 80.21 | 10.24 | 40.51 | 33.16 | 67.35 | 23.51 | 98.3 | 2.1 |
| **Load based with noise** | 79.56 | 12.31 | 44.37 | 34.22 | 61.23 | 29.3 | 99.31 | 3.13 |
| **Share memory with noise** | 75.43 | 13.57 | 50.12 | 32.51 | 62.91 | 28.47 | 98.82 | 2.29 |
| **IP timing channel with noise** | | | 45.66 | 20.45 | 75.66 | 23.66 | 98.33 | 5.15 |
| **Flush-reload with noise** | 73.21 | 13.81 | 41.28 | 32.67 | 66.95 | 22.71 | 99.12 | 4.1 |

**Table2. performance on memory detector and Hypercall detector**

| Local channel | Without detector | With Hypercall detector | With memory detector | Overload(%) with Hypercall detector | Overload(%) with memory detector |
|---|---|---|---|---|---|
| Cache based | 13550.6(ms) | 19106.35(ms) | 18293.31(ms) | 34 | 31 |
| Load based | 600.5(s) | 858.715(s) | 876.73(s) | 43 | 41 |
| Share memory | 325.5(s) | 481.74(s) | 465.465(s) | 45 | 42 |

experiments.

As shown in Table 1, under normal conditions, except the variance, the others had a high success rate and low FP (false positive). However, on the noise condition, the two-phase, variance and ϵ-similarity had significant differences in the success rate and FP. Thus, the noise from the normal processes was still a key factor in the impact of false positives. From the experimental results, our method maintained a high success rate, more than 98%, even under noise conditions, which means our method is more stable and the timing channel behavioral characteristics were accurately summarized.

### 6.4 Performance

Currently, placing a hook on the kernel to obtain the hypercalls log is the main input of the timing channel detection methods. Our method first proposed the memory activities records as the basic input. Thus, our main concern was the different performance impacts between them.

To ensure the accuracy of the detection, we reduced the impact on the timing channel to ensure its normal operation and to avoid timing channel antidetection. We conducted three controlled experiments to observe the impact of the detector on the performance of the timing channel. Except for the two-phase, which was the hypercalls detector, the others used the memory detector, including the method introduced from the network channel detection. For the network packets monitoring, the existing research was used to intercept the network card was implemented in the same way as our method, so it was not compared.

Our prototype system runs on the operating system, which means if the system is busier, it will cause a certain overload using real-time monitoring. As shown in Table 2, the load-based and share memory timing channel have more CPU consumption, and the overload is higher than the cache-based channel. The average increase was 10%. The overload difference was lower between the hypercall detector and the memory detector. The memory detector was only lower than the hypercall detector by approximately 3%, which is acceptable for users and does not affect the stability of the guest VM.

## 7. CONCLUSION AND FUTURE WORK

We tested four typical cross-VM timing channels in IaaS and obtained a satisfactory result. However, the upgraded versions of these channels were not tested, nor were other types of timing channels such as memory bus-based, hard drive, etc. Therefore, more long-term signatures of these timing channels must be summarized to improve the identification algorithm.

The memory activities records and packet records are a good basis for further research. More research on network timing channel detection is recommended. Some of the analysis methods used for the network packets can also be used for the memory activity. In this paper, the variance and ϵ-similarity method from the network channel both had a good result.

The basic idea of our method applies to many kinds of virtual machine systems. Next, complete and stable systems can be implemented on other virtualization platforms such as VMware VBOX, etc.

In the paper, we investigated timing channels in IaaS, summarized the behavior signatures of these timing channels, and then proposed a method to identify and investigate timing channels based on the signature. We designed a complete set of forensics steps and implemented our prototype system on Xen. The experimental result showed that the prototype successfully identifies these timing channels, even under the disturbances from normal processes.